\documentclass[aps,prb,preprint,groupedaddress]{revtex4}
\usepackage{graphicx}
\usepackage{xcolor}
\usepackage{float}

\usepackage{amsmath}
\usepackage{bm}
\usepackage{amsfonts}

\begin{document}

\title{Non-existence of the s-f volume-collapse transition in  solid gadolinium at pressure}

\author{Qingchen Li, Hossein Ehteshami, Keith Munro, Miriam Marqu\'es, Malcolm I McMahon and Graeme J. Ackland}
\email[]{g.j.ackland@ed.ac.uk }

\affiliation{ $^1$Centre for Science at Extreme Conditions and School of Physics and Astronomy, University of Edinburgh, Edinburgh, U.K. EH9 3FD}

\begin{abstract}

Gadolinium has long been believed to undergo a high pressure phase transition with a volume
collapse around 5\%. 
Theoretical explanations have focused on the idea of electrons transferring from the extended $s$-orbital to the compact $f$-orbital. However, experimental measurement has been
unable to detect any associated change in the magnetic properties of the  $f-$electrons\cite{fabbris2013different}.
Here we resolve this discrepancy by showing that there is no significant volume collapse, beyond what is typical in high pressure phase transformations.
We present density functional theory calculations of
solid gadolinium under high pressure using a range of methods, and revisit the experimental situation using X-ray diffraction (XRD).  The standard lanthanide
pressure-transformation sequence involving different stackings of close-packed planes: $hcp\rightarrow 9R\rightarrow
dhcp\rightarrow fcc\rightarrow d-fcc $ is reproduced.
The so-called ``volume collapsed'' high-pressure phase is shown to be an unusual stacking of close-packed planes, with  $Fddd$ symmetry and a density
change less than 2\%.  The distorted {\it fcc (d-fcc)} structure is revealed to arise as a consequence of antiferromagnetism.  The
theoretical results are shown to be remarkably robust to various treatments of the
$f$-electrons. The key result is that there is no XRD evidence for volume collapse in Gadolinium. The sequence of phase transitions is well described by standard DFT.  There is no need for special treatment of the $f$-electrons or evidence of $f$-electron bonding.  Noting previous spectroscopic evidence is that there is no change in the $f$-electrons we conclude that high pressure Gadolinium has no complicated $f$-electron physics such as Mott-Hubbard, Kondo or valence transition.

\end{abstract}
\date{\today} 
\maketitle 

\section{Introduction}
Over the last 30 years, calculations of high pressure phases using
density functional theory (DFT) and the Kohn-Sham Hamiltonian have shown a
remarkable level of agreement with experiment.  This occurs despite
the very simple quasi-local approximations used for the
exchange-correlation interactions, such as the local density (LDA) and
generalized gradient approximations (GGA).  Crystal structure prediction\cite{pickard2011ab,wang2012calypso} is
a particularly forgiving problem for DFT, because there is no explicit
dependence of the exchange correlation energy on the atomic positions\cite{hohenberg1964inhomogeneous,kohn1965self,perdew1983physical,perdew1996generalized}.

A notable exception to the rule has been so-called ``strongly
correlated'' systems, typically those where integer numbers of
electrons are localised on particular atoms.  For example, iron oxides
are wrongly predicted to be metallic with LDA/GGA, a situation which
required fixes such as adding some arbitrary amount of Hartree-Fock
exchange\cite{perdew1996rationale}, or a Hubbard $U$ term\cite{shick1999implementation} to force the $d$-electrons to remain
localised below the Fermi energy\cite{hubbard1963electron,meng2016density}.
It has generally been assumed that the problems encountered with
$d$-electrons would be even more severe in materials with
$f$-electrons, such as lanthanides.

The lanthanides (Ce to Lu) are characterised by the increasing number of
$4f$ electrons which, following Hund's rules, adopt the highest possible
value of spin.  The high stability of the filled spin-up $f-$shell means
that both Eu and Gd have 7 $f-$electrons with spin 7/2. 
At ambient pressure, the $f$-electron states lie far from the Fermi
level\cite{sabiryanov1997bulk,kurz2002magnetism} and the bonding is dominated by the $s$ and $d$ bands: this gives rise to a
common pressure-induced phase transition sequence comprising
different stackings of close-packed layers: hcp (space group
$P63/mmc$, hP2 in Pearson notation, and $h$ in close-packed stacking
notation\cite{loach2017stacking}) $\rightarrow 9R$ ($R\overline{3}m$, hR3, $hhf$)
$\rightarrow$ double-hcp ($P63/mmc$ hP4, $hf$) $\rightarrow$ fcc
($Fm\overline{3}m$ cF4, $f$).  The close-packed stacking notation
reveals a monotonic tendency away from hcp-like $ABA$ stackings toward
$ABC$-type stacking with pressure, a trend which is also observed with
reducing atomic number, to the extent that hcp does not occur in Sm
except at high temperature, and 9R is also absent from Ce to Pm.  
At elevated temperatures, the bcc structure may be observed.
At still higher pressure a distorted-fcc (dfcc) $(R \overline{3}m$ and hR24) 
structure appears, and beyond that low-symmetry post-dfcc phases, typically reported as being monoclinic.  A volume change of typically 5\% but as large as 9\%\cite{chesnutthesis2001} has been reported at the transitions to the post-dfcc phase.

In previous diffraction experiments on Gd\cite{GROSSHANS1992,xiong2014strength,akella1988high,hua1998theoretical,errandonea2007structural,chesnutthesis2001,montgomery2016high}, not all transformations were identified, e.g. missing hcp \cite{akella1988high}, missing fcc\cite{hua1998theoretical}, 
missing post-dfcc \cite{akella1988high}; the systematically-absent (106) reflection was observed in a phase reported as 9R \cite{montgomery2016high}; the stability ranges of the different phases have been somewhat different; the dfcc phase has been identified as hexagonal \cite{akella1988high}, monoclinic \cite{chesnutthesis2001}, and rhombohedral \cite{errandonea2007structural}; and the
post-{\it d-fcc} phase, which was not observed by Akella {\it et al} \cite{akella1988high}, was reported as I-centred monoclinic by Hua {\it et al} and Chesnut, but with volume changes of 5\% and 9.4\%, respectively, 
\cite{hua1998theoretical,chesnutthesis2001} but C-centred monoclinic by Errandonea {\it et al} with a volume change of 5\% \cite{errandonea2007structural}.

The similarity between all lanthanides suggests that the number of $f-$electrons
does not play a central role in determining the stacking sequence.
However, at much higher pressures all the lanthanides were believed to undergo a
``volume-collapse'' transition to a denser but non-close packed structure.  
An obvious cause for such a
transition would be the transfer of electrons from delocalised $s$ or $sd$ states
into the tightly-bound $f-$state\cite{ackland2003two,maddox20064,yoo2008electronic,samudrala2014structural,lim2014origin,sarvestani2014effect} - exactly the type of process which DFT
describes badly\cite{hua1998theoretical}. 

The three most popular theoretical explanations for such a
pressure-induced collapse are
\begin{itemize}
\item the $s-f$ valence transition
model\cite{ramirez1971theory}, in which a conduction band electron is transferred
into the $f-$band causing a reduction in the ionic radius; 

\item the Mott-Hubbard
model,\cite{johansson1974alpha} where the 4$f$ states undergo a
local-to-itinerant transition, leading to a significant contribution
to crystalline binding

\item the Kondo volume collapse
model\cite{allen1982kondo}, where the localized 4$f$ level approaches
the Fermi energy giving a sharp increase in the Kondo temperature. 
\end{itemize}

The
$4f$ electrons play a critical role in all these theoretical explanations.  However, experimental X-ray absorption and emission spectroscopy \cite{fabbris2013different,yoo2008electronic,maddox20064}
showed that the $f$-electrons were largely unaffected by the transformation.  In this paper we resolve this contradiction in favor of the experiment.

\section{Quasi-closed Packed Phases at Pressure}

\begin{figure}
\centering
\includegraphics[width=0.8\columnwidth]{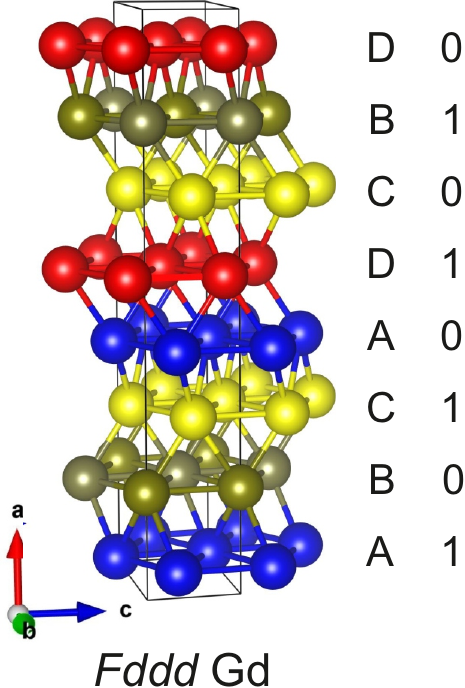}
\caption{The $Fddd$ structure as found in Gd, with the layers labelled according to the ABCD and 01 notations}
\label{fig:Fddd}
\end{figure}

Our recent high pressure X-ray
data led to a revised picture for the collapsed phases across the lanthanides\cite{mcmahon2019structure}.  
An entirely
new family of crystal structures was revealed: based on quasi-close packed
layers where each successive layer has atoms  situated above the mid-point between two
atoms in the preceding layer, resulting in 10-fold coordination.  By
analogy with the ABC notation for close packing the observed
structures were labelled ABCADCBD (Tb) ABCD (Pu) ABC (Sm), and the
possibility of other, short-repeat, stackings noted.  These structures
were shown to be a better fit to the X-ray data than the previously-assumed
body-centred monoclinic, $C2/m$
 or mC4 structures\cite{hua1998theoretical,errandonea2007structural,husband2012europium}.
In addition to having no layer repeats (e.g. AA is forbidden), we note that
these stacking structures all satisfy an additional condition that each layer
is different from either of the TWO below it, (ABA is forbidden).
This allows a concise notation to uniquely categorize the structures, labelling each layer by
\begin{itemize} 
\item 0 = same as three layers below
\item 1 = different from three layers below 
\end{itemize} 
In this notation, the three observed structures are the simplest
(ABCADCBD$\rightarrow 10$, ABCD$\rightarrow 1$, ABC$\rightarrow 0$),
whereas shorter sequences become longer (eg ABCDB$\rightarrow  11110)$.
This notation is illustrated for the observed $Fddd$ Gd (10) structure in Fig.\ref{fig:Fddd}.

Using a combination of the $hf$ notation for close packing, which explains the sequence\cite{loach2017stacking}, and this 01
notation for the new family, we can comprehensively study the possible
structures adopted by Gd.  A previous study included
electronic structure calculations using Dynamical Mean-Field Theory
(DFT+DMFT) to focus on the band structure and the $f-$electrons.  The
DMFT enables a detailed treatment of the unscreened moments, which
were believed to be {\it ``key for the correct description of the structure at high pressure.''}\cite{mcmahon2019structure}.  This belief arose because
a volume collapse due to  $s-f$ transfer obviously requires a
correct treatment of the $f-$electrons.  However, the DMFT method was found to
have instabilities as shown by the thermodynamically-impossible
discontinuity of Gibbs free energy in the DMFT-calculated
enthalpy-pressure relation\cite{mcmahon2019structure}.

\begin{figure}
\centering
\includegraphics[width=\columnwidth]{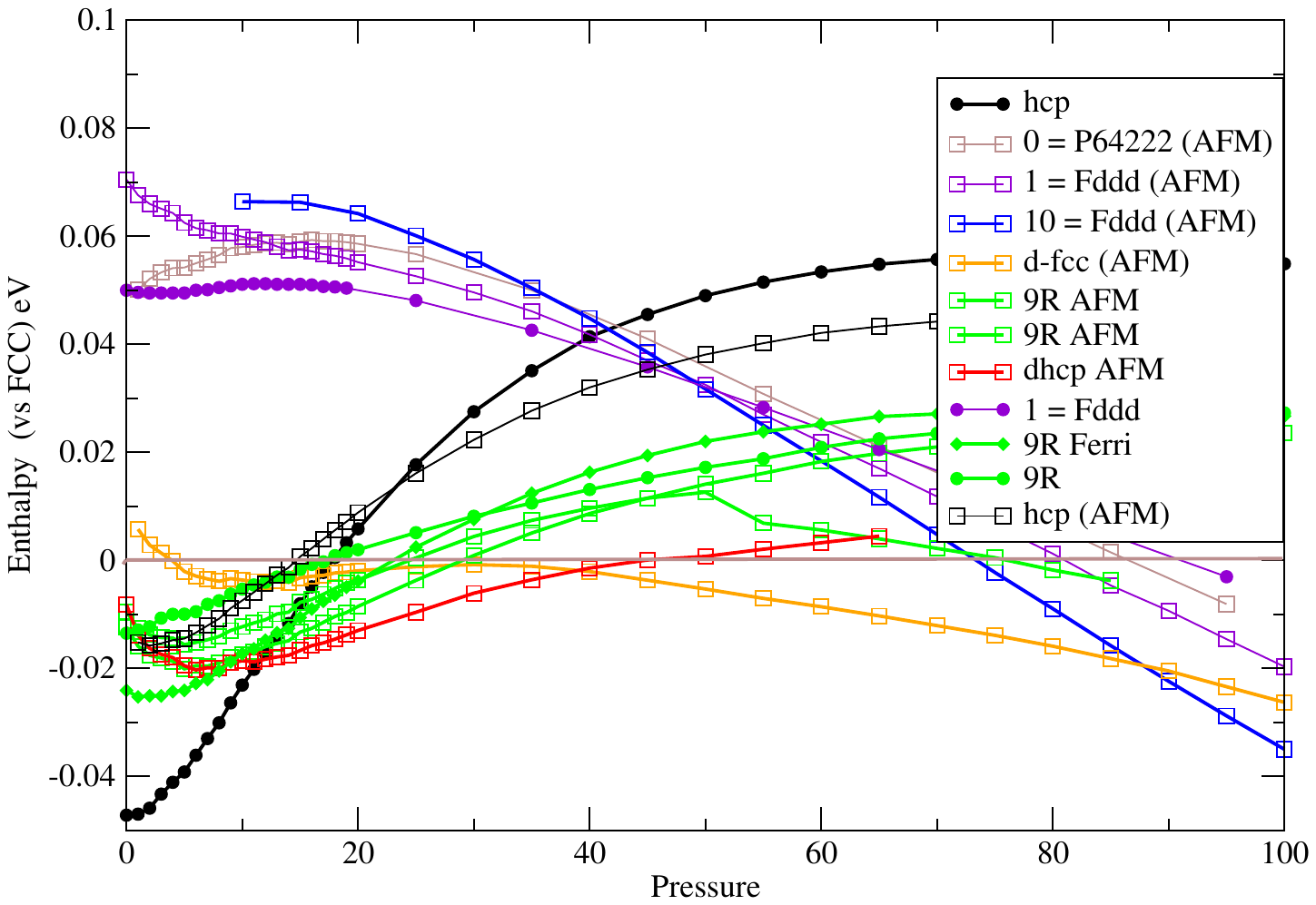}
\caption{ 0-K calculated enthalpy for various phases relative
  to ferromagnetic $fcc$. Filled circles show ferromagnetic phases, open ones AFM.  The implied sequence of phase transitions is
  $hcp$ (black) -$dhcp$ (red) - $fcc$ (brown) -$Fddd$ (blue).  The 9R phase (green, see Supplemental materials for AFM definitions) has a slope intermediate
  between $hcp$ and $dhcp$, but is only metastable in these calculations. Antiferromagnetism breaks $fcc$ symmetry, so AFM-$fcc$ has a small distortion from $fcc$, referred to as $d-fcc$.  At higher temperatures the paramagnetic 9R and $fcc$ also have regions of stability.}
\label{enthalpy}
\end{figure}

\section{Determining the necessary level of electronic structure theory}
\subsection{Philosophy and methods}
In this paper,  rather than using the most sophisticated treatment available, we ask how simple can a DFT calculation be, while still obtaining the correct series of phase transformations.  This approach will demonstrate which aspects of the physics are essential to understanding the sequence.  We choose to examine gadolinium, because its half-filled $f$-shell is likely to be most amenable to a simple treatment.

We used the CASTEP code with ultrasoft pseudopotentials and PBE\cite{segall2002first,clark2005first,vanderbilt1990soft,perdew1996generalized}
functional.  We treated the $5s5p4f6s5d$ and above electrons as valence,
giving 18 explicitly calculated electron states per atom.  We used a
plane wave basis set with a 425\,eV cutoff.  We also considered use of a
Hubbard-$U$ which will have the effect to localize, but not split, the
$f-$electrons. The Hubbard-$U$ is generally regarded as the minimal plausible theory
for $f-$electrons, and we used throughout the standard value of $U=6.7$\,eV \cite{oroszlany2015magnetism}
giving a good fit to the experimentally determined electronic structure. 
The exception is where we test the sensitivity to U.

{ In all cases we checked k-point convergence to less than 0.1meV using grids up to 20,000 k-points.  All structures were geometry-optimised to a target pressure with respect to cell parameters, angles, magnetic moment and internal coordinates.  This relaxation is essential to determine pressures and densities. The close-packed structures all have c/a ratios close to the hard-sphere value, with the layer spacing slightly below ($\sqrt{2/3}$) of the in-plane neighbour distance. Examples of the input data, including  antiferromagnetic orderings, are given in Supplemental materials.  }  

\subsection{Importance of magnetism}
The simplest level of theory, non-magnetic calculations, results in an
$f-$band at the Fermi surface.  This is clearly incorrect, so all calculations presented here allow for spin dependence\cite{segall2002first}.

At ambient pressure Gd has a ferromagnetic hcp structure with a Curie
temperature of 293\,K\cite{lide2005magnetic}, with some susceptibility
anomalies\cite{kaul2000gadolinium,coey1999gadolinium}.  On further
increase in pressure the situation is unclear, but may be similar to
Sm which exhibits layered antiferromagnetic structures at pressure
with a Neel temperature around 60K\cite{vohra2016magnetic}.  Consequently, in our calculations we consider both ferromagnetic and various antiferromagnetic versions of each structure. 

Figure.\ref{enthalpy} shows that using the standard value of $U$ gives
the typical lanthanide sequence of phase transformations, with the exception that 9R is slightly too high in energy.  An example of the
density of states is shown in Figure \ref{DoS}.  The $f-$electrons are split
into a spin-up and spin-down bands, well above and below the Fermi
energy.  By projecting the wavefunctions onto localised orbitals, we find that the valence states have hybrid $s-d$ character.
There is no splitting within the $f-$electrons levels of the
same-spin, in contrast to the DMFT result\cite{mcmahon2019structure}. 
We can therefore conclude
that {\it the sequence of crystal structures does not depend on the
  accurate treatment of $f$-electrons}.   Most remarkably, this observation
also applies to the volume-collapsed phase.

It is interesting to note the similar shape of the majority and minority spin bands in hcp, aside from a shift which places the Fermi level in a density of states minimum. It suggests that, in addition to the localised $f$-magnetization, the $hcp$  structure exhibits Stoner-type $sd$-ferromagnetism\cite{stoner1948mechanism}. 
  
To investigate further, we varied the value of $U$, giving a
progressively poorer treatment of the $f$-electrons.  Even more
remarkably, this {\it still} does not affect the sequence of phase
transformations, even when the $U$ is absent altogether.

\subsection{Effect of Hubbard U}
{ In our calculations we used PBE+U with U=6.7eV to "force" the $f$-electrons to localise.  At issue is whether the calculated Kohn-Sham orbital represents a particular electron confined within the cell - in which case it should not feel its own electrostatic potential - or if it represents a sum of fractions of many delocalised electrons which do repel each other.  Self interaction correction calculations demonstrate that localizing only $f$ electrons is energetically favorable\cite{hughes2007lanthanide}, and the Hubbard-U is a convenient way to capture this effect.}

In Fig.\ref{DoS-U} we show the effect of changing U on the density of states, calculated for the AFM-Fddd phase at 90\,GPa but typical of all structures.
The very sharp peak corresponding to the half-filled
  $f-$band lies well below E$_F$, and the main effect of +$U$ is to shift
  this peak. This treatment means that the peak is not split,
  and the figures show that it also does not broaden, hybridize or
  contribute to the valence band.
 
  Regardless of the choice of $U$, the $f-$band does move closer to E$_F$
  with pressure, and this is essentially unaffected by the crystal structure (see Supplemental materials for calculation with U=0).   
  
{ Even with U=0eV, } the occupied $f$-states lie below the 
  $sd$-band and the unoccupied $f$-states do not hybridise, appearing as a distinct but broader peak.   Consequently, the value of $U$ has no significant effect  on the energy differences between phases and the phase
  transformation sequence. 
 
  Interestingly, the $f$-band remains localised even with $U=0$, and the use of the Hubbard $U$ is not essential to describe the sequence of transitions.
  In fact, the densities calculated with $U=0$ are in better agreement with the experiment than with the $U=6.7$\,eV value deduced from first principles\cite{oroszlany2015magnetism}.  The volume change at the so-called "volume collapse" is smaller than the volume change for the ferromagnetic $hcp$ to antiferromagnetic $dhcp$ transformation.  
  
In early work using non-selfconsistent lattice parameters taken from experiment, 
  and an unrelaxed c/a ratio, the minority f-band was found close to the Fermi level\cite{kurz2002magnetism,hughes2007lanthanide,heinemann1994lsda,heinemann1994magnetic}.  The ground state was found  to be antiferromagnetic: the correct  ferromagnetic ground state was recovered with the addition of a Hubbard U\cite{kurz2002magnetism} or self-interaction correction\cite{hughes2007lanthanide} to ensure that the f-electrons are localised.   Surprisingly, our fully relaxed calculation shows LDA and PBE both have a ferromagnetic ground state at zero pressure even without these corrections. The incorrect previous prediction must therefore be due to something other than treatment of exchange.
  Since U has the effect of reducing the density (or, for fixed volume, reducing the pressure), it seem likely that the unrelaxed calculations fell into either high stress or high-pressure regime where hcp antiferromagnetism is long-known to be favored\cite{heinemann1994magnetic}.
  
  With U=0, at P=0, and fully relaxed $hcp$, we find that LDA favors the FM state over AFM by 10meV/atom, whereas gradient corrected PBE favors FM by 22meV/atom.  

\section{Role of Ferromagnetism, Antiferromagnetism and Paramagnetism}

In general, antiferromagnetic states are favoured at high pressure, because the antialigned spins mean atoms are not kept apart by Fermi repulsion.  Paramagnetic states are favoured at high temperature because of the entropy of disordered spins.  A recent review shows that the subtle magnetic ordering in lanthanides can be explained using indirect exchange (Ruderman–Kittel–Kasuya–Yosida) built on density of states and equation of states calculated using standard DFT \cite{gimaev2021magnetic}.  In the case of Gd, recent work shows that at low pressure Gd is a ferromagnet, with significant $sd$-contributions to the moment, up to 7GPa, when the bulk magnetisation vanishes \cite{tokita2004rkky,mcwhan1965effect}. {The "volume collapse" regime occurs at an order of magnitude higher pressure than the loss of moment, but the atomic moments are still intact: therefore the high pressure structures must be either antiferromagnetic or paramagnetic.

The calculations allow us to make some observations about the nature of the structures. At 0\,GPa, all the close packed structures were found to be ferromagnetic within their range of stability, except for $fcc$. Above 45\,GPa ferromagnetic $fcc$ becomes unstable with respect to antiferromagnetic ordering.   

Antiferromagnetic ordering is frustrated in all structures considered here.  
For layered structures, it is always possible to find an antiferromagnetic ordering which preserves the observed crystal symmetry.
However, $fcc$ is unique among the considered phases, it is impossible to decorate the
$fcc$ lattice with spins while maintaining the cubic symmetry, and on
relaxing the atomic positions in the antiferromagnetic supercell they
move away from $fcc$.  Thus any antiferromagnetic order on the $fcc$ lattice necessarily breaks cubic symmetry.
  No such constraint exists for paramagnetism, so above the Neel temperature fcc and local antiferromagnetic interacions become compatible.
This provides an elegant explanation for the
transition to the d-$fcc$ phase: it is the structural signature
of a magnetic phase transition, and the distortion vanishes at higher temperature.}

{\color{black}
According to Hund's rule, the spin on each atom from the f-electron is maximised at $7/2$.  Our calculations find that there is also an additional, pressure-induced contribution from the $s$ and $d$ electrons (see Fig 2b).
Such magnetic symmetry breaking in metals can be approximately described by any fixed-spin antiferromagnetic lattice model, such as Ising, Potts or Heisenberg.
Any such model implies that there is a paramagnetic
phase at elevated temperature, with higher symmetry.  This provides a simple explanation for the {\it d-fcc} phase: the small distortions to $fcc$ come from antiferromagnetic order and disappear in the paramagnetic phase.}

\begin{figure}
\centering
\includegraphics[width=0.32\columnwidth]{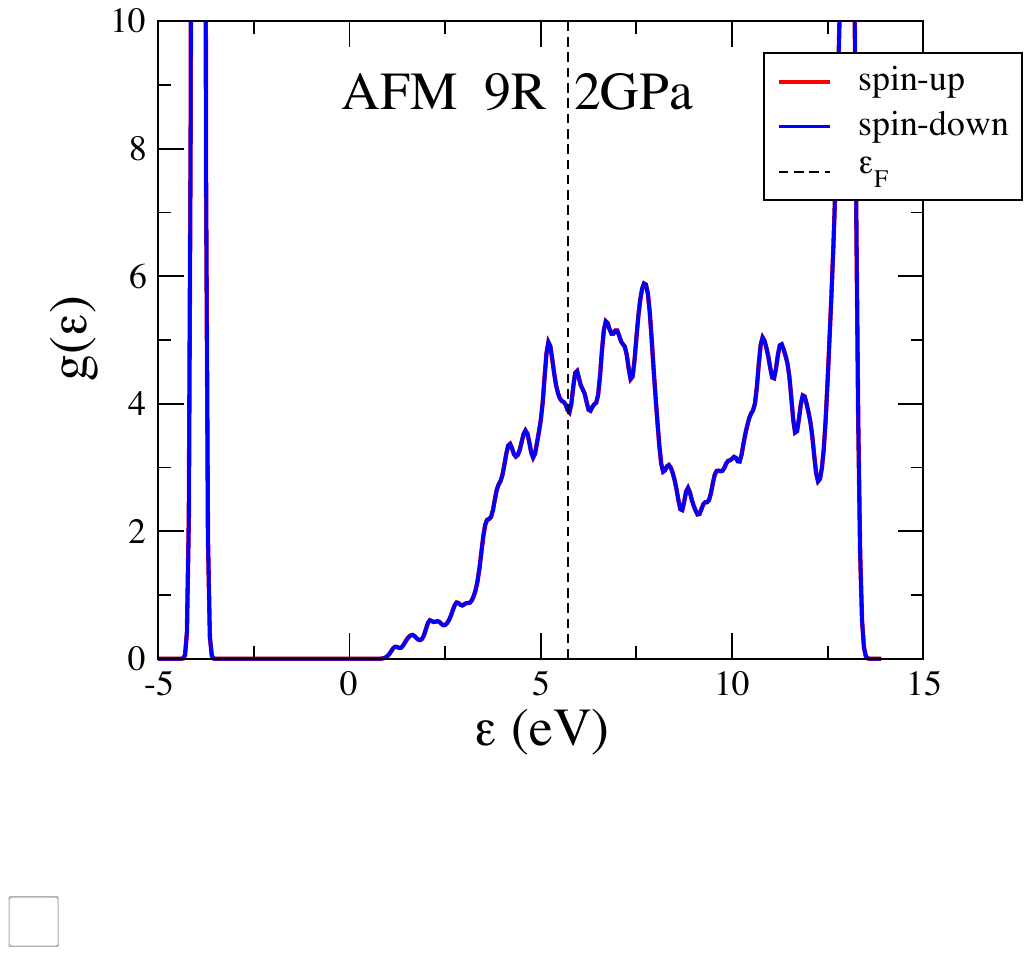}
\includegraphics[width=0.32\columnwidth]{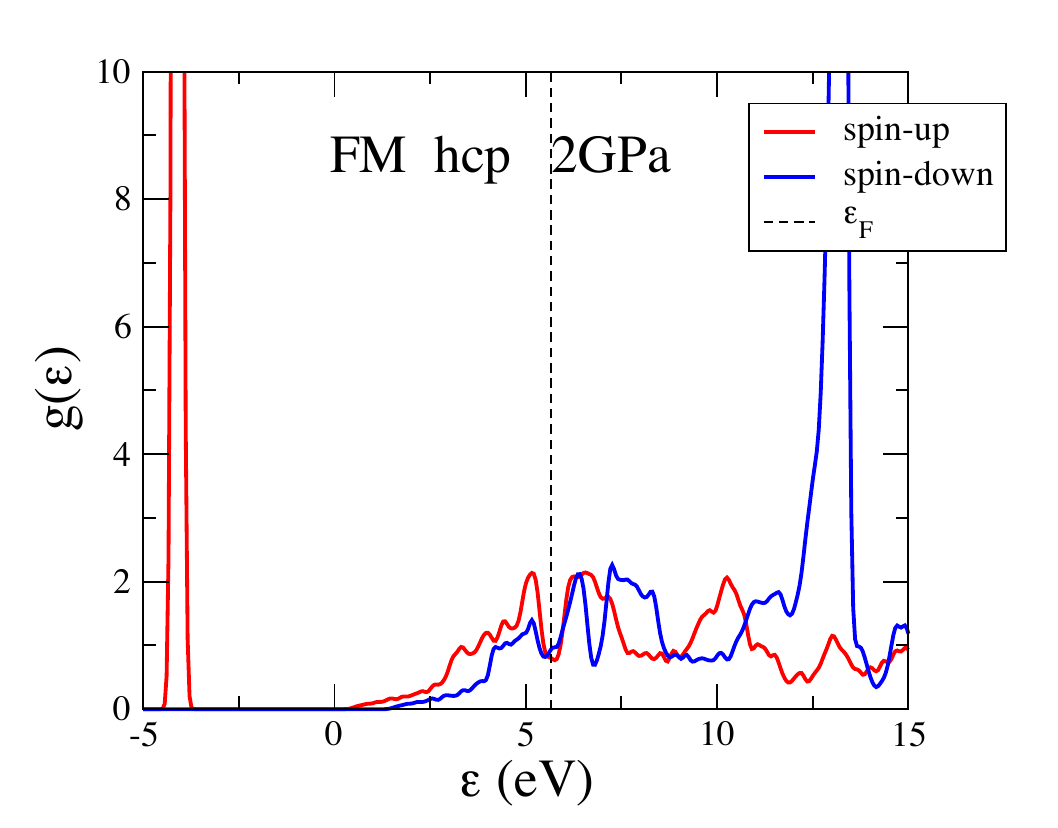}
\includegraphics[width=0.32\columnwidth]{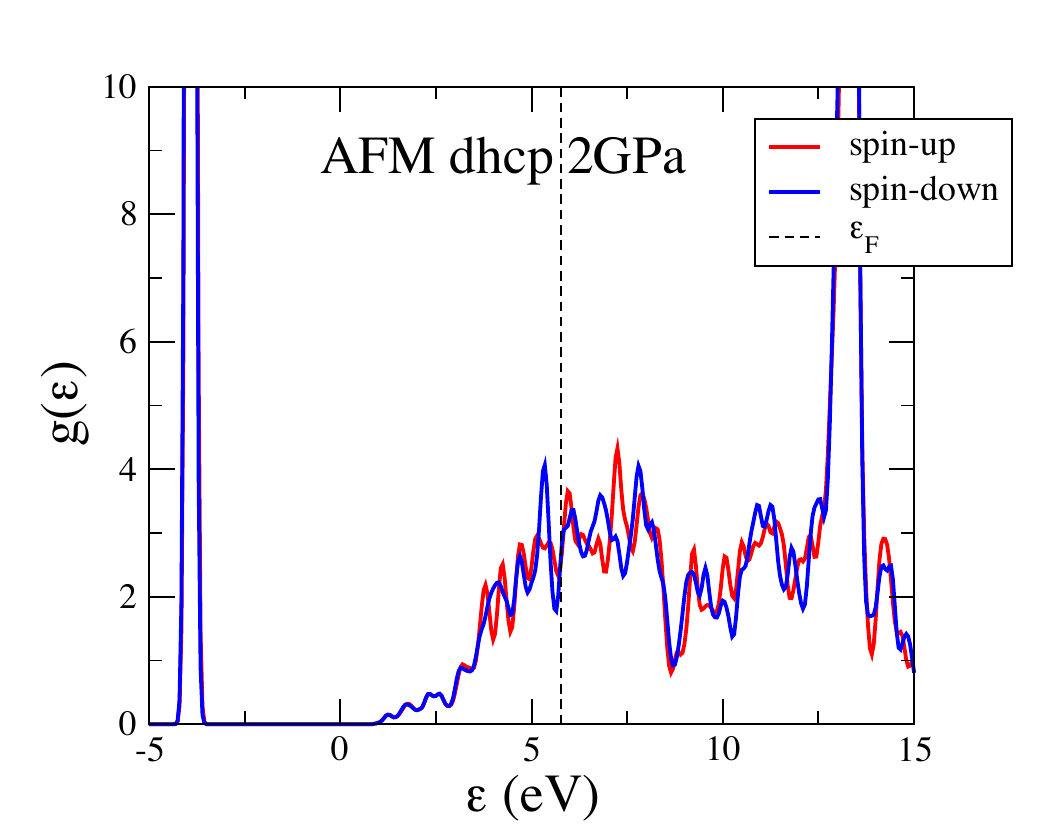}
\caption{ Calculated Density of States per unit cell for three competing structures at 2GPa.
(a) antiferromagnetic 9R (b) Ferromagnetic $hcp$ (c) antiferromagnetic $dhcp$. The very sharp peak corresponding to the half-filled $f-$band lies well below E$_F$, with the unfilled $f-$band well above  E$_F$.  States around the Fermi level have hybrid $s-d$ character.
\label{DoS}}
\end{figure}

\begin{figure}[h]
\centering
\includegraphics[width=0.95\columnwidth]{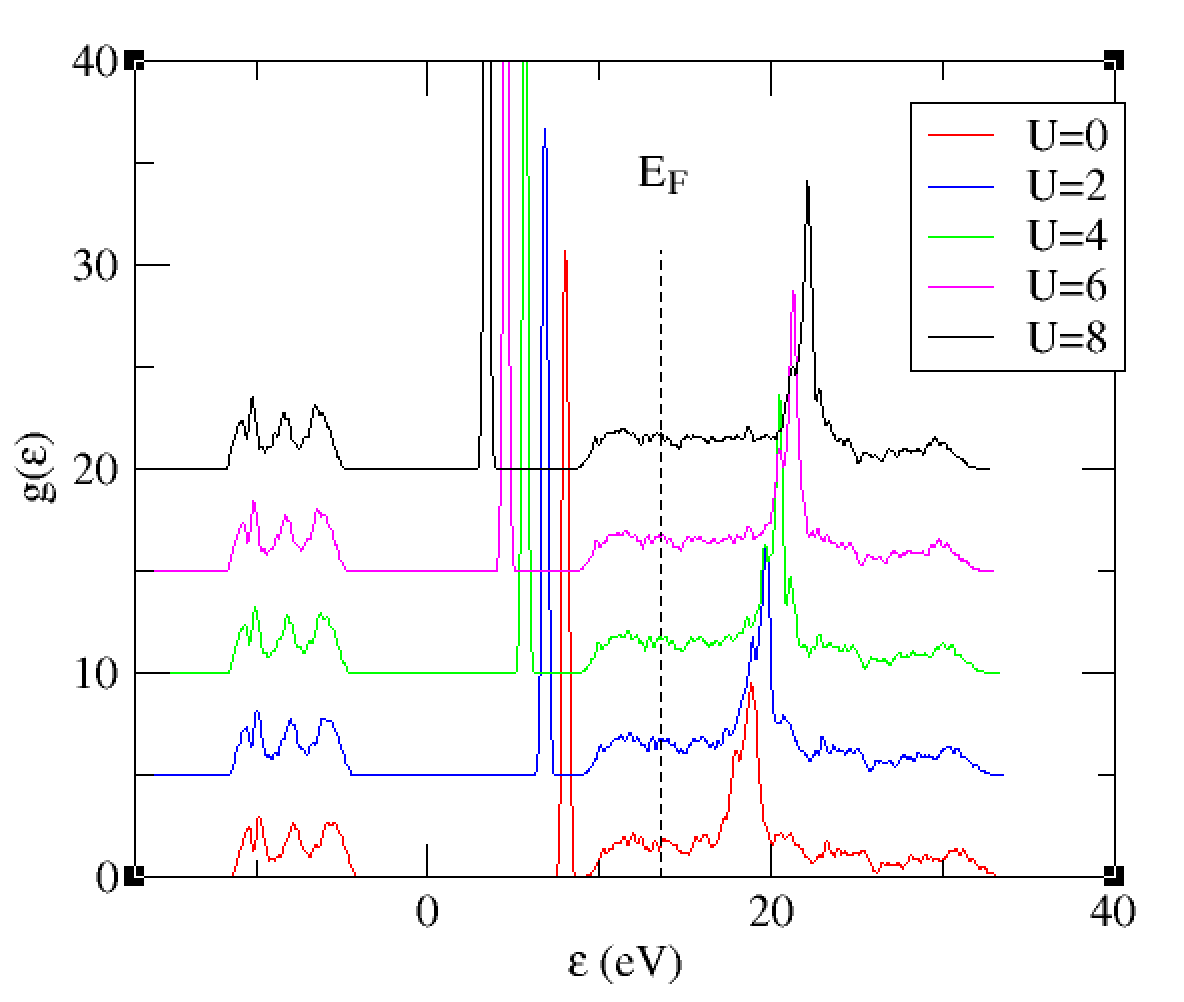}
\caption{ Calculated Density of States for AFM-Fddd phase at 90 GPa  with various values of U as shown in eV.  {\color{green} Associated changes in density and c/a are given in the supplemental materials}
\label{DoS-U}}
\end{figure}

\section{Free energy calculations}

{ In addition to the enthalpy calculated by DFT, contribution to the free energy  come from the phonons  and associated zero point energy (section \ref{sec:phon}).
For two different crystal structures, the free energy difference can be written as a sum of three parts

\begin{equation} \Delta G = \Delta \mathcal{G}^{mag} + \Delta G^{NM} + \Delta G^{phon} \end{equation}

In this way of partitioning the energy, the DFT enthalpy difference $\Delta H^{DFT}$ calculated from the relaxed structures includes both magnetic ground state and non-magnetic contributions. 
}

\subsection{Phonon free energy calculation \label{sec:phon}}

We carried out phonon free energy calculations in the harmonic approximation using CASTEP.  Harmonic phonon frequencies are calculated using the as-implemented finite displacement lattice dynamics  method\cite{warren1996ab,ackland1997practical,clark2005first}
At 0\,GPa, we compared the stable ferromagnetic hcp phase with the lowest energy (ferrimagnetic) 9R phase. This comprises a double close-packed layer of up-spin followed by a single layer of down-spin, resulting in a macroscopic moment: this arrangement is neither ferromagnetic not anti ferromagnetic, hence the slightly irregular use of the term Ferrimagnetic.
It is the lowest enthalpy decoration of spins we found, below ferromagnetic,  alternating close-packed layers, and alternating $[11\overline{2}0]$ lines within the close packed layers, the arrangement which maximises the number of opposite-spin pairs

\begin{figure}
\centering
\includegraphics[width=1.2\columnwidth]{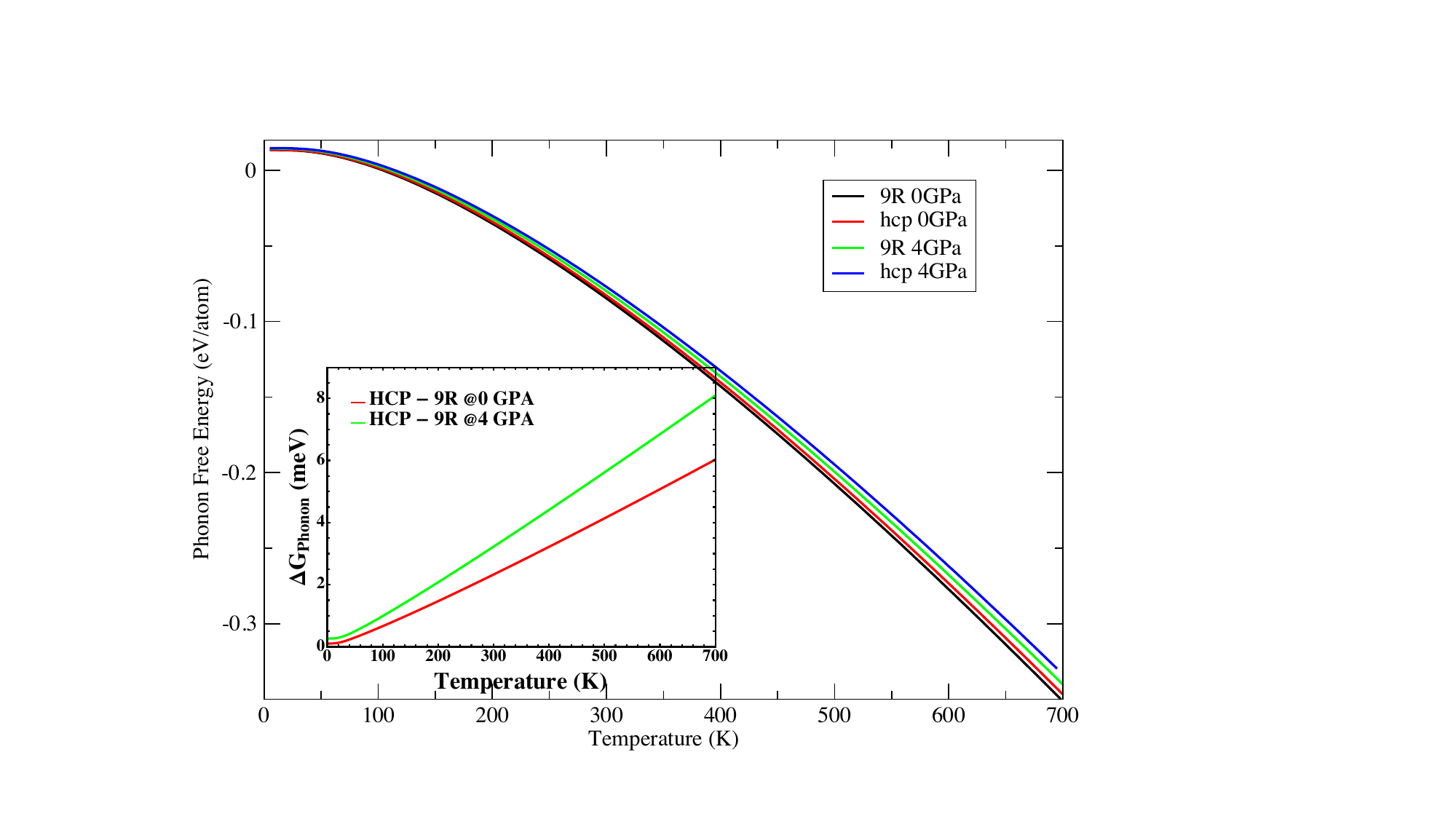}
\caption{  Harmonic contribution to Phonon Free Energy for 9R and $hcp$ at 0 and 4GPa.  Inset shows the difference between phases at either 0GPa (red) or 4GPa (green) is of order 1meV, favoring 9R, an order of magnitude smaller than the magnetic effect
\label{phon-free}}
\end{figure}

\begin{figure}
\centering
\includegraphics[width=\columnwidth]{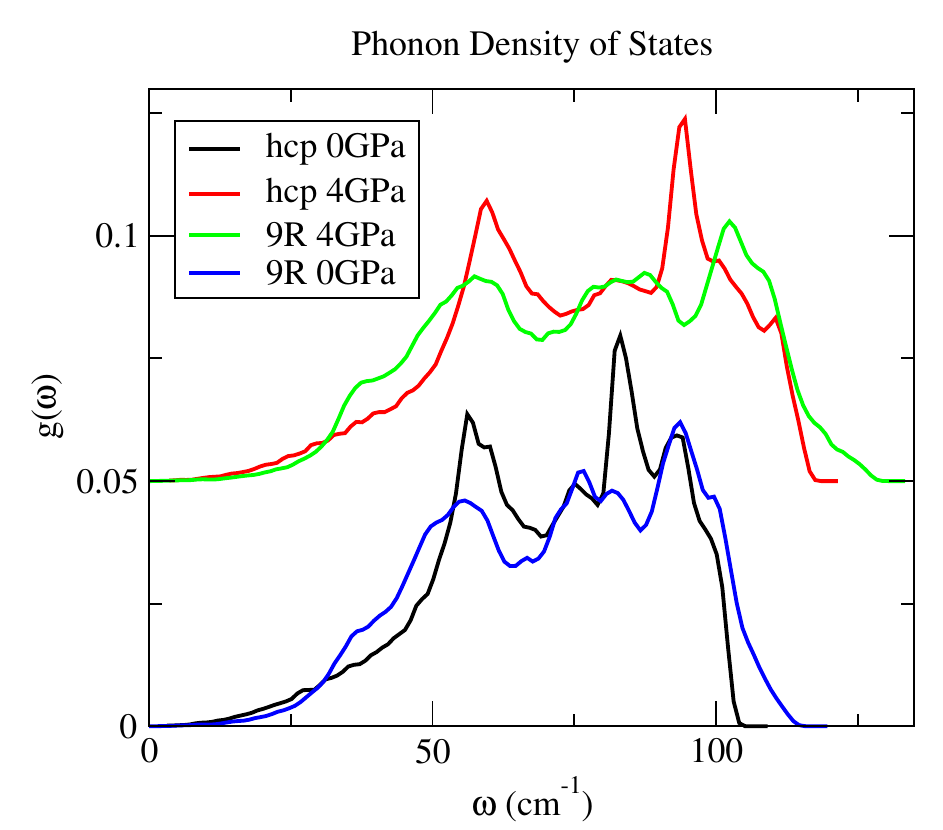}
\caption{ Phonon densities of states for 9R and $hcp$ at 0\,GPa and 4\,GPa (displaced).  The zero point energies are 0.01354eV/atom and 0.01450\,eV/atom for 9R;   0.01365\,eV/atom and 0.01470\,eV/atom for $hcp$ at 0 and 4\,GPa respectively.
\label{phonDos9R}}
\end{figure}

Figure. \ref{phon-free} shows the variation in the phonon contribution to the free energy in the harmonic approximation.  The main feature to note is that the $hcp$ and 9R phases are extremely close, e.g. at 300\,K, 0\,GPa $hcp$ is -83\,meV and 9R is -85\,meV, the 2\,meV difference is therefore an order of magnitude lower than the magnetic effects.  The phonon densities of state themselves are shown in Figure. \ref{phonDos9R}

The main feature of the phonon free energy is that, although absolute values are large, the differences between phases are two orders of magnitude smaller - a few meV.  The difference in zero-point energy is another order of magnitude smaller.   Thus the phonon free energy has negligible effect on phase stability.

\subsection{Magnetic free energy calculation}

{\it 
The enthalpy associated with antiferromagnetic ordering is sufficient to describe all the high pressure DFT results consistently with experiment, even if magnetic entropy is ignored.    At low pressures the
experimentally-reported 9R phase\footnote{Recall that there is no published XRD pattern to support this report} has very slightly higher DFT enthalpy than hcp, so we investigate whether magnetic entropy could explain the discrepancy. The idea is that hcp has a Curie temperature around room temperature, so that if 9R has a Neel/Curie temperature significantly below that, it will have higher entropy, which could stabilize it.  
}

To estimate the magnetic ordering free energy, we calculated the free energy using a spin model on an close-packed lattice with near-neighbor interaction.  
Typically, the DFT calculation gives a spin of around S=4 on each atom, 3.5 of which comes from the $f$-electrons and the other $\approx 0.5$, reducing with pressure, comes from the $sd$-band. For an Ising or Heisenberg model write the magnetic energy as

\begin{equation} \label{eq:H}
\mathcal{H}^{mag} = -J \sum\limits_{\left\langle i,j \right\rangle^{\prime}} \sigma_i \sigma_j.
\end{equation}
$J$ can be either positive (ferromagnetic, ground state energy $-6J$) or negative (frustrated antiferromagnetic, ground state energy $2J$).  Henceforth, we will drop the $mag$ superscript and use script symbols for the contributions which depends on the magnetic orientation. For each close-packed crystal structure we set $|\sigma|=1$ and calculate $J$ from the difference between ferromagnetic and antiferromagnetic structures. At zero kelvin, ferromagnetic  $\mathcal{G}=-6J$ and antiferromagnetic $\mathcal{G}=sJ$, so that 


\begin{equation} 
8J = H_{DFT}^{FM}-H_{DFT}^{AFM} \label{eq:8J}
\end{equation}

 The non-magnetic part is assumed to be temperature independent, and is calculated indirectly from.  

\begin{equation} \Delta H^{NM} = H^{DFT}_{9R} - H^{DFT}_{hcp} - \mathcal{G}^{9R}(T=0) + \mathcal{G}^{hcp}(T=0)
\end{equation}

  These energies are included in the DFT calculations, so avoiding double counting leads to:

\begin{equation} \Delta G = \Delta \mathcal{G} + \Delta H^{DFT}
+ 6\Delta J +  \Delta G^{phon} \end{equation}

with $\Delta$ indicating the difference between hcp and 9R.

We use the colinear-spin model both for simplicity and because our DFT calculation are based on two-band (Ising) spins. Free energy has calculated with other models (Heisenberg, Potts) using EMFT which gives some indication of the uncertainty introduced by the choice of the  Ising approximation. 
We  calculate these
 estimated temperature effects and using the EMFT approach assuming Ising spins\cite{ehteshami2020phase,honmura1979contribution}, and validated the EMFT against Monte Carlo simulations\cite{ackland2006magnetically}.  Further details are given in Supplementary Materials.

\subsection{Parameterization of EMFT}

The parameters $J$ were fitted to the DFT values for the difference in enthalpy between FM and AFM structures.  Consequently, $J$ takes a different value for each crystal structure.    Where more than one AFM structure was considered, $J$ was fitted to the lower value.  We note that the enthalpy difference includes the $P\Delta V$ term which arises from the volume difference between FM and AFM.  The negative thermal expansion of Gd arises from the fact that as spins flip thermally in the FM-hcp phase, the reduced exchange interaction allows for compression.

At low pressures, $|J|$ is lower in 9R than in hcp, so the Curie Temperature in 9R is well below that in hcp.  The Ising model overestimates $T_c$ - using the quantum Heisenberg model gives a lower value, and the classical Heisenburg model lower still.  

  Even in the Ising case, this estimated value of $T_c$ is very sensitive to assumptions: a calculation at the experimental density gives a value some $20\%$ lower than the experimental one; fitting a second neighbour model gives a similar reduction\cite{graham1965magnetic,vorobev1966crystal,zi9z1972influence}. 
  
  For 9R, there is ambiguity about the AFM structure: The lowest energy structure we found among the many possible AFM 9R structures is ferrimagnetic, with the (two) $h$ and (one) $f$ layers having opposite spin.  All the AFM structures considered lie within 10\,meV per atom.  $dhcp$ is also antiferromagnetic, with a similarly low Neel temperature.   
  
  We also note that the AFM configurations are typically 0.5\% denser than the ferromagnetic ones  due to fewer exchange repulsions. At pressure this contributes to the DFT enthalpy. This results in the negative linear thermal expansion coefficient in the $c$-direction.

In the current work we are interested in phase transformations with pressure, and therefore free energy.  But we note that  
magnetic transition temperatures can be calculated using EMFT.   For a Heisenberg model Plascak~\cite{Plascak08phase} states that the Ne{\'e}l and Curie temperatures are comparable to the Ising case, assuming the same coupling constant $J$.  However they assume that the Ising Spin is $S=\frac{1}{2}$ while the Heisenburg is $S=\sqrt(\frac{1}{2}.\frac{3}{2})$. which, in fitting equation \ref{eq:H}, introduces an additional  factor of 3.

\begin{figure}
\centering
\includegraphics[width=\columnwidth]{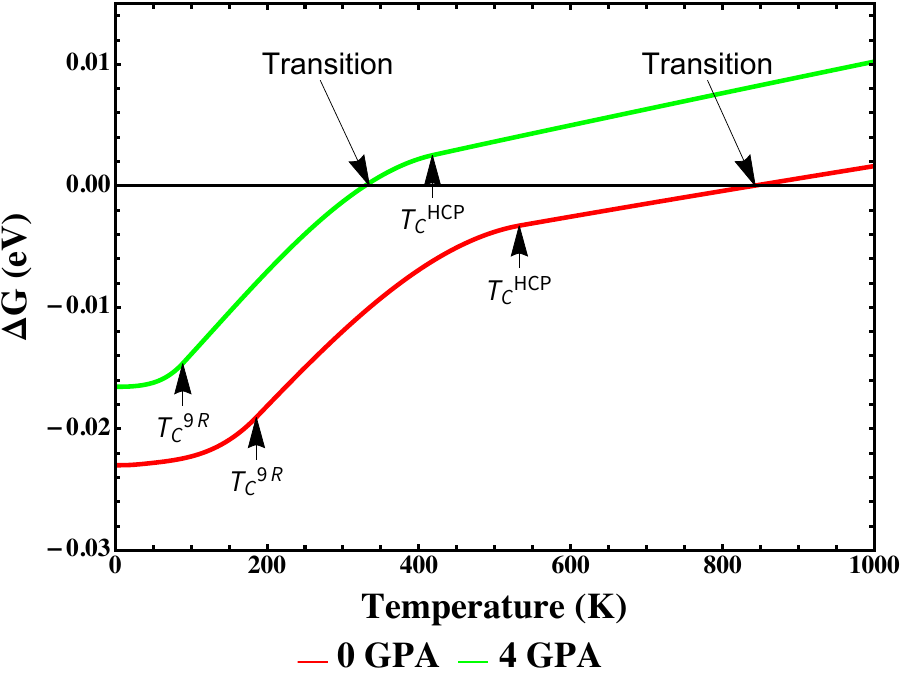}
\caption{Free Energy differences between hcp and 9R phases at 0GPa and 4GPa as a function of temperature. The hcp-9R transition temperatures ($\Delta G=0$) are shown, along with the Curie temperatures for each phase/pressure.
The main temperature dependent contibution comes from magnetism,  $\Delta\mathcal{H}$. This is estimated here using effective mean field theory with  hcp  8J= 33.5 (0GPa)  26.3 (4GPa) and  9R  J=11.7,  5.6  meV.
The phonon free energy difference, including zero-point is only about 0.002\,eV, stabilizing 9R slightly.
\label{fig-EMFT}}
\end{figure}

{ We also consider the effect of using a Potts or Heisenberg model in the EMFT approach\cite{ehteshami2020phase,ehteshami2021high}, again with the coupling constants fitted to AFM-FM differences from the DFT.  The general picture is the same, the 9R phase has a significantly lower paramagnetic transition temperature than hcp acquiring extra entropy from the magnetic degrees of freedom.  The transition temperature can change by a factor of up to 3 depending on whether the magnitude of the Heisenberg model is equal to the Ising Spin, or to $\sqrt{\sigma(\sigma + 1)}$, or the limiting case of a classical moment.  By analogy with fcc\cite{Plascak08phase} with $s=4$ the Heisenberg result for Tc is a factor of 3 lower, so the two models bracket the experimental value.
Regardless of Tc, the stabilization of 9R by paramagnetism at room temperature is a robust result. 

 Figure \ref{fig-EMFT} shows the effect of adding all contributions to the  free energy.  We see that magnetic effect increase the stability of 9R.  Pressure also stabilizes 9R, such that is becomes stable at room temperatures and modest pressures.  Exact values of the transition P and T are sensitive to the choice of model, but the physical effect is robust, ultimately depending only on |J| being lower for 9R, and the energy scale for $T_c$ being a similar order of magnitude to room temperature
 }
\section{X-ray diffraction evidence}

Previous experimental evidence for ``volume collapse" transitions in the lanthanides has been based on X-ray diffraction (XRD) measurements. However, XRD is not a direct measure of density: it is necessary to know the correct crystal structure before it is possible to determine the density. We have also reexamined the experimental situation in Gd.  Room temperature angle-dispersive x-ray diffraction experiments were performed on the high-pressure I15 beamline at the Diamond Light Source using diamond anvil cells and a MAR345 detector. The Gd samples were cut from samples of 99.99\% purity and two contaminants phases were identified in the diffraction patterns at low pressures, each of which could be fitted with an $fcc$ structure and then eliminated from the Rietveld refinement of diffraction profiles from the pure Gd. The full experimental details are published elsewhere \cite{munro2017high,mcmahon2019structure}.

The as-loaded sample showed a number of peaks which could not be accounted for by the $hcp$ phase, and were assigned to the 9R phase. The transformation to $dhcp$ was first detected at 8.9\,GPa and found to be complete by 10.2\,GPa. The $dhcp$ phase transformed directly into the d-$fcc$ phase at 33.6\,GPa, by-passing the fcc phase completely, similar to the results reported by Hua {\it et al.} \cite{hua1998theoretical} 
We found that the previously-assumed monoclinic $mC4$ structure is incorrect  \cite{mcmahon2019structure}. The ``volume collapsed" phase is now identified as orthorhombic, space group $Fddd$ and with 16 atoms/cell, which enabled us to reexamine the equation of state of Gd to 87 GPa.  Fig.\ref{xrdfig} shows a waterfall plot of diffraction data across the d-$fcc\rightarrow Fddd$ transformation, and compares the implied experimental equation of state with the DFT calculation.
 
\begin{figure}
\centering \includegraphics[width=\columnwidth]{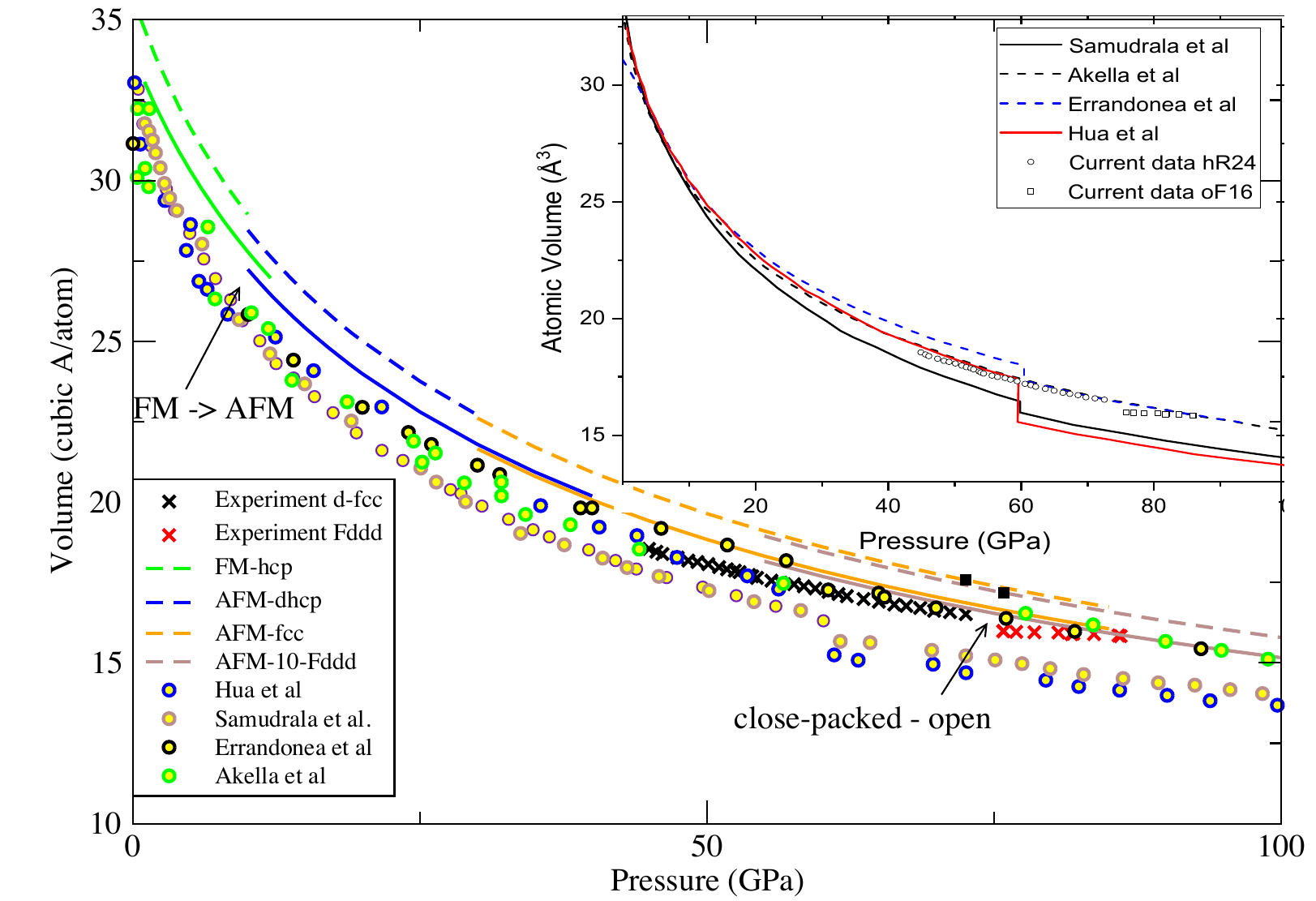} 
\caption{ Equation of state for the various phases $hcp-dhcp-fcc-dfcc-Fddd$ comparing calculations (lines) with data points from previous experimental  EoS\cite{errandonea2007structural,akella1988high,samudrala2014structural,hua1998theoretical} (circles) and our data including $Fddd$ (crosses). The experimental $V_0$ value is 33.10 \AA$^3$/atom whereas the calculated  is 33.85 or 35.67\,\AA$^3$/atom with U=0 (solid lines) or $U=$\,6.7eV (dashed lines).  Black squares emphasize the calculated volume change.   (lower) Experimental data only, with  our data shown as symbols and previous work shown as smooth fits to previous data up to the transition, and to reported densities (assuming incorrect structures) above the transition.  \label{EoS}
}
\end{figure}

\begin{figure}
\includegraphics[width=\columnwidth]{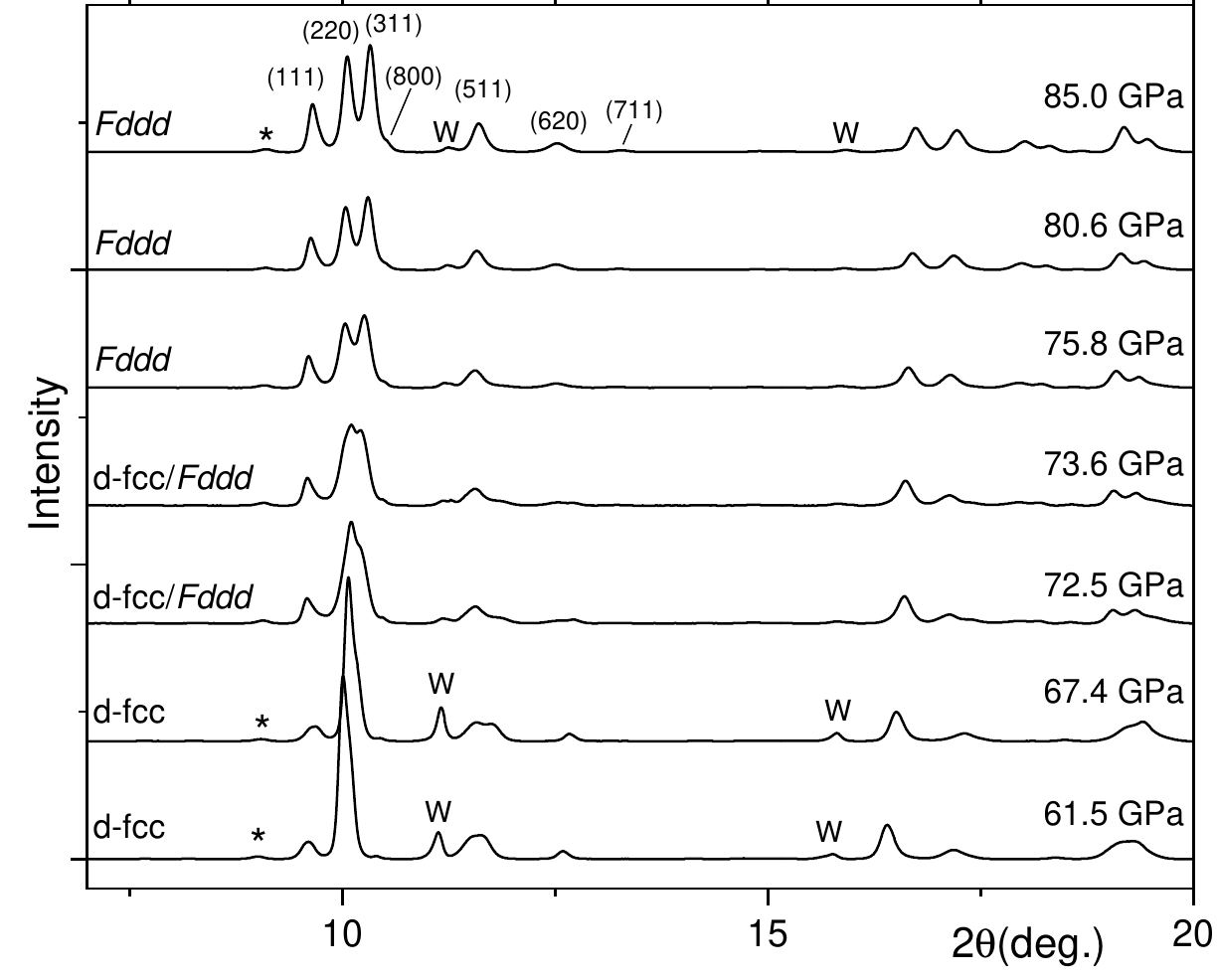} 
\caption{Waterfall plot of integrated diffraction profiles obtained from Gd between 61 and 85\,GPa, illustrating the d-$fcc\rightarrow Fddd$ transition at $\sim$73\,GPa. The low-angle peaks in the $Fddd$ phase at 85 GPa are indexed according to Pearson notation oF16. Peaks from the tungsten gasket and from a minority contaminant phase are identified with `W' and asterisks, respectively. \label{xrdfig}}
\end{figure}

\begin{figure}
\centering
\includegraphics[width=0.43\columnwidth]{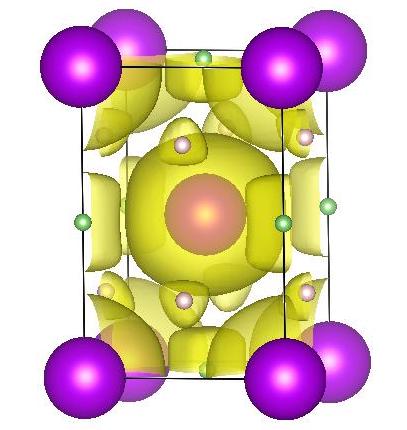}
\includegraphics[width=0.43\columnwidth]{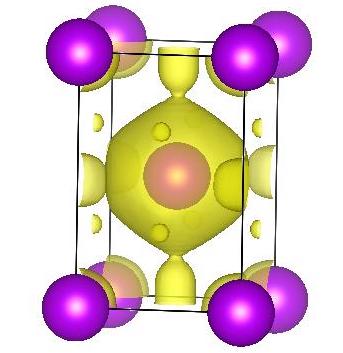}
\includegraphics[width=0.43\columnwidth]{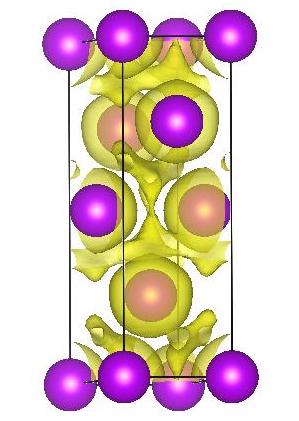}
\includegraphics[width=0.43\columnwidth]{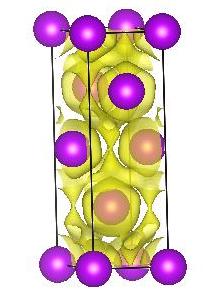}
\caption{Electron Localization function isosurface  (ELF=0.3)  at 85\,GPa for d-$fcc$ (a) majority band (b) minority band and for $Fddd$  (c) majority band (d) treating $f$-electrons as core.  
The image we get for both d-$fcc$ and $Fddd$ is  quite similar,
most of the volume is occupied by the ELF-basins associated with the Gd atoms. At\,85 GPa, each Gd contributes with approx. 1.3 electrons to the valence basins, and at 125\,GPa, 1 electron.
The valence maxima show low values of ELF, particularly for the  $Fddd$
structure. The ELF value at these maxima also decreases with pressure.
The low value of the ELF in regions outwith of the basins associated with the atoms is also clear looking at two planes containing the valence maxima. Gd atoms appear as purple spheres and 2$b$ and 4$d$ valence maxima for the d-fcc structure as green and pink spheres, respectively.
\label{ELF}}
\end{figure}

 \section{How big is the volume change?}

 DFT calculations show the $mC4$ structure to be very high in energy, but the 10 stacking ($Fddd$) agrees well with the experiment, being more stable than 0 ($P6_422$)  and 1 (also $Fddd$) stackings.   

The ``collapsed'' $Fddd$ phase becomes stable over d-$fcc$ above 85\,GPa. 
In the 75\,GPa DFT calculation, the d-fcc phase has a volume 16.69\,\AA$^3$/atom
compared with $Fddd$ 16.53\,\AA$^3$/atom, 
rising to 16.65  \AA$^3$/atom if Fddd is forced to be ferromagnetic.   This is similar to the experimental value which drops by 1.2\% of $V_0$, from 16.35\,\AA$^3$/atom to 15.98\AA\,$^3$/atom (Fig.\ref{EoS}).  This calculated difference in density between d-$fcc$ and $Fddd$ is smaller than the transition from $hcp$ and $dhcp$, so there is no reason to call it a "volume collapse".

The density at which DFT predicts the transformation is in good agreement with the experiment, but the pressure is slightly higher than the experiment (Fig.\ref{EoS}): this discrepancy is typical for DFT.  The uncertainty in experimental transition pressure is striking: previous experiment put it at around 60\,GPa, our experiment sees the transition at 73\,GPa, while the calculation gives 87\,GPa, albeit at zero Kelvin.

\section{Nature of the electronic structure}

In the simple metals, and the trivalent elements Sc and Al, the onset of complex high pressure phases has been attributed to electride transitions and Fermi-Surface Brillouin interactions. We tested whether this is important in Gd by looking more deeply at the electronic structure of $Fddd$ and $fcc$.

\subsection{ELF calculations}
Electrides show localised charge at non-nuclear interstitial positions which is best investigated using the electron localization function for DFT  (ELF). 

Figure \ref{ELF} shows the ELF=0.3 isosurface for d-fcc and $Fddd$ structures. For d-$fcc$ two types of ELF attractors appear on $2b$ and $4d$ sites with low values of ELF, and the localization window - the difference between the ELF at the bond point connecting them and the value of the ELF at the attractors - decreases with pressure. So, the system becomes more free-electron-like with pressure since a small delocalization window is linked to high electron delocalization.  In $Fddd$, there are more inequivalent attractor sites, but the picture is similar - the small localization window is typical of a metallic system with no sign of non-nuclear localized electrons and the material is not an electride.
We recalculated these structures and calculating the topology using a pseudopotential treating the $f$-electrons as core. The qualitative image is very similar with ELF maxima located at the same sites, further evidence that the $f$-electrons do not participate in the bonding.

{\color{black}
\subsection{Fermiology}
The Fermi-Surface - nesting concept has been invoked as an explanation for the transformation to complex structure in simple metals\cite{ackland2004origin,degtyareva2006simple}.  In order to lower the free energy, the nesting must break symmetry and
manifest as a dip in the electronic density of states at the Fermi level.  This also creates a cluster of diffraction peaks at 2k$_F$. There is no evidence of this in any DOS (Fig \ref{DoS}) of diffraction pattern.  Thus we conclude that the $Fddd$ structure is not stabilized by interaction between the Fermi Surface and the Brillouin zone.

$Fddd$ has multiply folded bands in a tiny Brilluoin zone, but effect of pressure in the $fcc$ is easier to see thanks to the monatomic unit cell (supplemental materials).  At 0GPa $fcc$ has parabolic s-bands similar to a free-electron metal, but by 75GPa the energy of the zone boundary states has dropped significantly with energy minima at X and L points
}

\section{Discussion}

In sum, we have presented a series of DFT calculations of high pressure phases in gadolinium.  These have shown that, despite the difficulty of describing localised $f$-states in DFT, the correct sequence of structures can be obtained using standard methods.  This indicates that the $f$-electrons do not play a role in the high pressure phase stability of Gadolinium.

Treating the phases as antiferromagnetic is important, and both $sd$ and $f$ bands exhibit magnetization. The symmetry breaking of the distorted-$fcc$ phase is shown to be a necessary consequence of the frustration of antiferromagnetism of that lattice.  All high-pressure phases apart for $hcp$ are predicted to be antiferromagnetic.
  
  The lanthanide sequence of close-packed stacking is broadly reproduced by DFT.  The status of 9R Gd is debatable: entropic effects were required to stabilise 9R over hcp in the calculation, and the XRD pattern showed non-hcp peaks but could not be Rietveld refined to 9R.  It may be that the pattern is the result of stacking faults in hcp, as seen in copper and lithium\cite{blackstock2001phase,ackland2017quantum}.
  
  DFT calculation even applies to the "volume-collapsed", phases which were previously associated with $s-f$ electron transfer.  Furthermore, they predict that there is no significant volume collapse - a finding borne out by the recent revision of the high-pressure crystal structure.  This work explains the failure of magnetic susceptibility experiments to verify any of the rival theories of $s-f$ electron transfer at volume collapse: neither exist.

We have concentrated here on gadolinium, and demonstrated that 
 $f$-electrons do not participate in the bonding, even at high pressure.
Other lanthanides have similar high-pressure structures to $Fddd$\cite{munro2017high}. It would be extraordinary if these structures, unique to the lanthanides,  were somehow stabilized by completely different physics.  So we expect that $f$-electrons do not participate in the bonding of any lanthanide below a megabar.

\begin{table}[h]
\begin{tabular}{|c|ccccc|}
\hline
$hcp\rightarrow$ &9R &dhcp&fcc&d-fcc&Fddd \\
$h$ & $hhf$ & $hf$ & $f$ &$f$ & 01 \\
\hline
Errandonea &2&6&26&33&60.5  \\
Samudrala &2&6.5&25&31&61\\
Montgomery & 2 & 6 & & &\\
Expt & 0 & 10 & - & 33.6 & 73 \\
Calc (FM) & - & 11 & 33 & - & 90\\
Calc (AFM) & - & 10 & 30 & 40 & 87 \\
\hline
\end{tabular}
 \caption{0\,K DFT transition pressures in GPa.  The calculated $hcp\rightarrow$9R transformation pressure is above $hcp\rightarrow dhcp$, so 9R has no predicted region of stability at 0\,K.
 \label{transpre}
 }
 \end{table}

\begin{acknowledgments} GJA and HE acknowledge the support of the European Research
Council Grant Hecate Reference No. 695527. GJA acknowledges a Royal Society Wolfson fellowship.
EPSRC funded studentships for KM, computing time (UKCP grant P022561). MIM is grateful to AWE for the award of a William Penney Fellowship. The computational work was supported by the ARCHER UK National Supercomputing Service. 
\end{acknowledgments}
\newpage\newpage\newpage\newpage\newpage\newpage\newpage\newpage

\bibliographystyle{apsrev}
\bibliography{Gd,Gd_mim}

\end{document}